\documentclass[prl,aps,twocolumn,showpacs,superscriptaddress,nofootinbib]{revtex4}
\usepackage{graphicx} 
\usepackage{dcolumn}  
\usepackage{bm}       
\usepackage{amsmath}
\usepackage{epsfig}

\def\ii{{\rm i}}

\begin{document}
\title{Graphene optical-to-thermal converter}
\author{Alejandro Manjavacas}
\email{alejandro.manjavacas@rice.edu}
\affiliation{Department of Physics and Astronomy and Laboratory for Nanophotonics, Rice University, Houston, Texas 77005, United States}
\author{Sukosin Thongrattanasiri}
\affiliation{Department of Physics, Kasetsart University, Bangkok 10900, Thailand}
\author{Jean-Jacques Greffet}
\affiliation{Laboratoire Charles Fabry, Institut d'Optique, Universit\'e Paris Sud, CNRS, 2 Av. Fresnel, 91127 Palaiseau Cedex, France}
\author{F. Javier Garc\'{\i}a de Abajo}
\email{javier.garciadeabajo@icfo.es}
\affiliation{ICFO-Institut de Ciencies Fotoniques, Mediterranean Technology Park, 08860 Castelldefels (Barcelona), Spain}
\affiliation{ICREA-Instituci\'o Catalana de Recerca i Estudis Avan\c{c}ats, Passeig Llu\'{\i}s Companys, 23, 08010 Barcelona, Spain}


\begin{abstract}
Infrared plasmons in doped graphene nanostructures produce large optical absorption that can be used for narrow-band thermal light emission at tunable frequencies that strongly depend on the doping charge. By virtue of Kirchhoff's law, thermal light emission is proportional to the absorption, thus resulting in narrow emission lines associated with the electrically controlled plasmons of heated graphene. Here we show that realistic designs of graphene plasmonic structures can release over 90\% of the emission through individual infrared lines with 1\% bandwidth. We examine anisotropic graphene structures in which efficient heating can be produced upon optical pumping tuned to a plasmonic absorption resonance situated in the blue region relative to the thermal emission. An incoherent thermal light converter is thus achieved. Our results open a radically different approach for designing tunable nanoscale infrared light sources.
\end{abstract}
\maketitle



The fabrication of intense, tunable infrared sources remains a challenge, despite the availability of quantum-cascade laser \cite{FCS94}, free-electron lasers \cite{UGK98}, and thermal emission bulbs. In particular, thermal emission has been extensively studied, leading to advances such as the control over its angular distribution from heated gratings \cite{GCJ02} and fundamental insights on the coherence properties of the associated electromagnetic field near the emitting surfaces \cite{CG99}. Additionally, thermal emission has been shown, in combination with resonant metamaterial designs, to be confined within infrared peaks of high quality factor --the ratio of the energy width to the central energy of a peak-- $Q\sim5$, which can be tuned by modifying the geometry \cite{LTS11}. Alternatively, fast electro-optical modulation can be achieved through the use of doped graphene, which has been demonstrated to display plasmons that are frequency-controlled via electrostatic gating \cite{JGH11,FAB11,paper196,FRA12,paper212,BJS13}, reaching $Q>50$ \cite{paper212}. By conveniently shaping the graphene, it is possible to reach plasmon-mediated complete optical absorption over large areas \cite{paper182}. Conversely, by virtue of Kirchhoff's law, black-body light emission occurs in heated graphene at frequencies corresponding to the full-absorption resonances. Recently, conversion of broadband light to quasi-monochromatic emission has been achieved \cite{DAM12}, while electrical modulation of the resulting emission has been successfully demonstrated \cite{BSS14}. Electrical modulation of the emissivity has been also predicted using quantum wells \cite{VMM14} and recently demonstrated at 600 kHz \cite{IDA14}.

\begin{figure}
\begin{center}
\includegraphics[width=80mm,angle=0,clip]{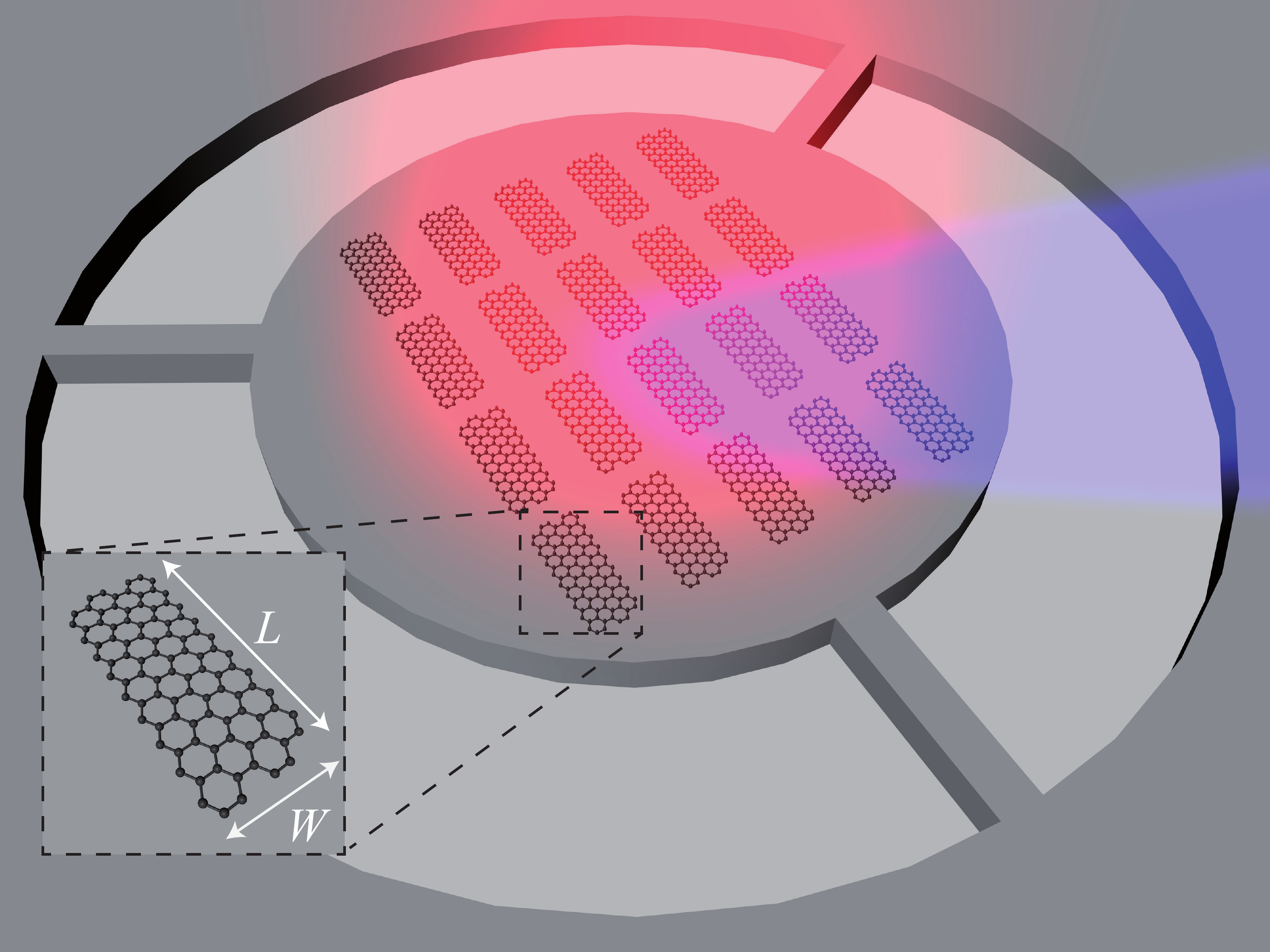}
\caption{Scheme of an optical-to-thermal converter device consisting of an array of graphene nanoantennas supported on a thermally isolated non-absorbing dielectric.} \label{Fig1}
\end{center}
\end{figure}

Here, we explore the possibility of combining electrically modulated resonant thermal emission from graphene with light pumping in order to realize an optical converter. The conversion efficiency is strongly dependent on the strength of the plasmon resonances relative to the background. Using realistic parameters to describe the optical response of graphene, we predict that $>90\%$ of the emission takes place within a single narrow plasmon line, with a net emission power $\sim10\,$W/m$^2$ under realistic heating conditions (e.g., as estimated from a grey body with the absorptivity of Fig. 3(b) below, at a temperature $T=600$\,K; incidentally, we use Gaussian units throughout this work). It should be noted that efficient narrow band emitter devices can be made based upon other types of optical resonators, such as for example plasmon nanocavities \cite{IMK08}, photonic crystals \cite{LMF03,CSJ06,IDA13}, metamaterials \cite{LTS11}, and semiconductors \cite{SLR13}. However, graphene is unique in its combination of electrical tunability, which allows one to actively manipulate its absorption and thermal emission wavelengths at much higher modulation frequencies than other materials \cite{IDA14}; low volume, which facilitates integration; and relatively low losses compared with other plasmonic materials.

We show in Fig.\ \ref{Fig1} a scheme of an optical-to-thermal converter. Graphene nanostructures are patterned on a thermally insulating dielectric membrane that is held in vacuum and mechanically sustained via narrow holding arms, so that thermal diffusion can be neglected and the main cooling mechanism of the graphene-substrate system is through radiative emission \cite{BBH14}. Heating can be introduced in different ways (e.g., Joule heating), but we concentrate here on optical pumping, realized at an absorption resonance frequency (e.g., tuned to a transverse-polarized plasmon of the graphene antennas depicted in Fig.\ \ref{Fig1}). The graphene islands are chosen to have a high aspect ratio, so that emission is enhanced at a lower frequency associated with a plasmon polarized along the longitudinal direction (see below). In what follows, we assume the substrate to be poorly absorbing in the frequency range under consideration, and thus, we neglect it in our analysis of the radiation emission. The final efficiency of the system critically depends on this assumption, which requires the use of poorly absorbing materials.

From Kirchhoff's law, the power emitted per unit frequency $\omega$ from a subwavelength structure at temperature $T_1$ held in vacuum is simply given by
\begin{equation}
P(\omega)=\frac{4}{3}\sum_{i=x,y}\sigma_i^{\rm abs}(\omega)\,E(\omega),
\label{Pw}
\end{equation}
where $\sigma_i^{\rm abs}(\omega)$ is the absorption cross-section for polarization along the principal direction $i$ (we neglect the response along the direction $z$ perpendicular to the graphene), $E(\omega)=(\hbar\omega^3/4\pi^2c^2)n_1(\omega)$ is the power radiated by a blackbody per unit frequency and unit area, $n_1(\omega)=[\exp(\hbar\omega/k_BT_1)-1]^{-1}$ is the Bose-Einstein distribution, and the $4/3$ factor correctly accounts for the angle-integrated emission from a dipole. Before analyzing an extended surface (e.g., the device of Fig.\ \ref{Fig1}), for which $\sigma^{\rm abs}$ is simply the absorptivity times the area, we discuss the properties of an individual antenna.

\begin{figure}
\begin{center}
\includegraphics[width=80mm,angle=0,clip]{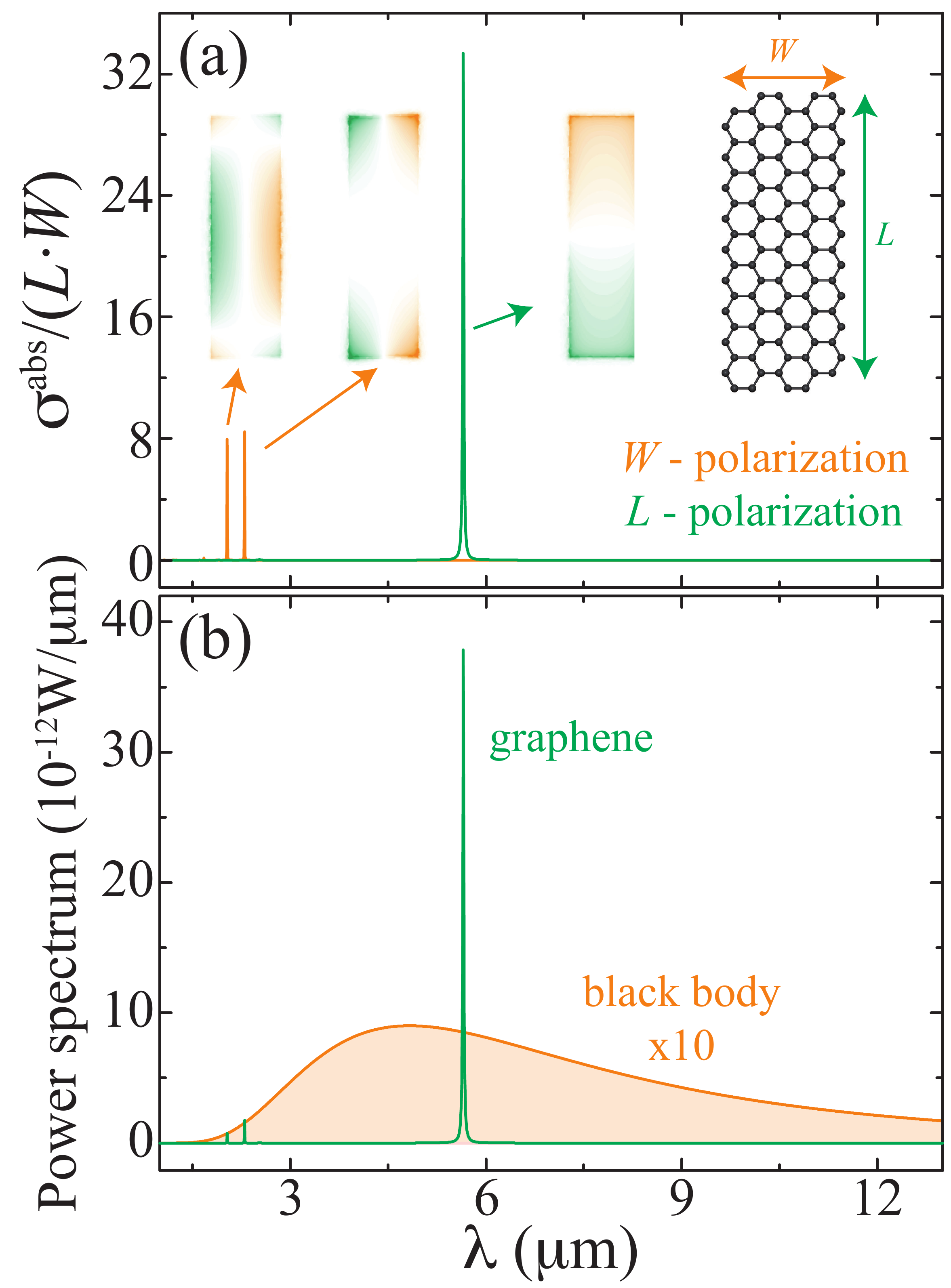}
\caption{{\bf (a)} Absorption of an individual rectangular graphene nanoantenna (length $L=60\,$nm, width $W=15\,$nm) for polarization along its width (W, orange line) or its length (L, green line). Induced density maps associated with the three observed dominant peaks are shown as insets. {\bf (b)} Thermal light emission of a single nanoantenna (emission power per unit wavelength), compared with the emission of a blackbody of the same area. The graphene is described through its local-RPA conductivity \cite{paper176} with Fermi energy, mobility, and temperature set to $1\,$eV, $10000\,$cm$^2/($V s$)$, and $600\,$K, respectively.} \label{Fig2}
\end{center}
\end{figure}

Figure\ \ref{Fig2}(a) shows a characteristic absorption cross-section of a self-standing graphene rectangle nanoantenna of dimensions $L\times W=60\,$nm$\times15\,$nm doped to a Fermi energy $E_F=1\,$eV, for incoming light polarized along the width (W, orange line) or the length (L, green line) of the nanostructure. Although the cross section is several times larger than the geometrical area of the nanoantenna for both of these polarizations under the assumption of a realistic mobility \cite{NGM04,NGM05} $10000\,$cm$^2/($V s$)$, the longitudinal mode produces four times larger absorption than the transversal ones. Interestingly, this coincides with the length-to-width ratio of the rectangle, as expected from a spectral sum rule based upon the dipolar character of the modes \cite{paper235}. The latter is confirmed by looking into the induced charge density maps [Fig.\ \ref{Fig2}(a), insets]. The thermal emission from this antenna is shown in Fig.\ \ref{Fig2}(b) at a temperature of 600\,K (green curve), as obtained from Eq.\ (\ref{Pw}). We emphasize that this emission is solely determined by the temperature and the absorption cross-section of the nanostructure. The high aspect ratio of the graphene island is important to have a large spectral separation between longitudinal and transversal modes, so that the former can then be placed closer to the blackbody emission maximum (orange curve), and thus, become the leading emission channel.

The three absorption resonances ($j=1-3$) observed in Fig.\ \ref{Fig2}(a) are spectrally isolated, so that the island can be described through a polarizability $\alpha_\omega$ near each of them (notice that the length $L$ is much smaller than the wavelength) with a Lorentzian profile \cite{paper235}
\begin{equation}
\alpha_\omega\approx L^3\frac{A_j}{2B_j/(1+\epsilon)-\ii\omega L/\sigma(\omega)}, \label{alpha}
\end{equation}
where $\sigma(\omega)$ is the graphene conductivity, whereas the coefficients $A_j$ and $B_j$ are only a function of the $L/W$ aspect ratio. More precisely, we find via finite-difference simulations (COMSOL) the values $(A_1,B_1)=(0.088,41)$ and $(A_2,B_2)=(0.12,54)$ for the transverse plasmons and $(A_3,B_3)=(0.20,6.4)$ for the longitudinal one. The effect of the substrate is included in this expression through its permittivity $\epsilon$.

\begin{figure}
\begin{center}
\includegraphics[width=80mm,angle=0,clip]{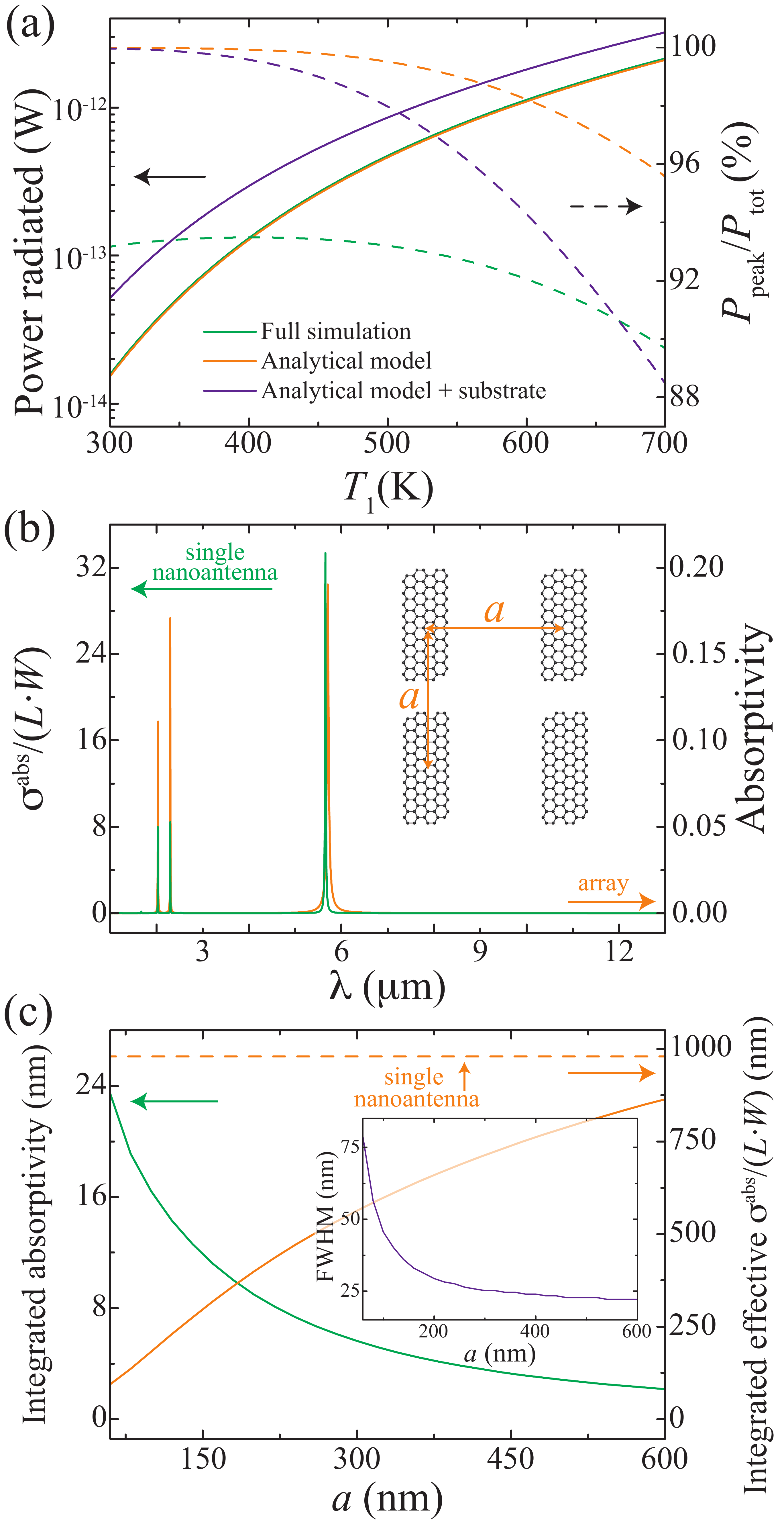}
\caption{{\bf Emission efficiency and emitted power.} {\bf (a)} We show the emitted power (left scale) and the fraction of emitted power in the main plasmon peak relative to the total emission (right scale) as a function of temperature for an individual graphene nanoisland as those considered in  Fig.\ \ref{Fig2}. Full simulations are compared with analytical theory with and without inclusion of a substrate of $\epsilon=2$ (see labels). {\bf (b)} Absorptivity of an extended graphene square array (right scale, period $a=120\,$nm) compared with the cross section of an antenna normalized to its area [left scale, taken from Fig.\ \ref{Fig2}(a)]. {\bf (c)} Integral over wavelengths around the main peak shown in (b) for the absorptivity (left scale) and the normalized effective absorption cross-section (right scale) of a square array as a function of spacing. The effective cross section is defined as the absorptivity times the unit cell area $a^2$. The integral for the normalized absorption cross-section of an individual island is shown as a dashed line for comparison. The inset shows the peak width as a function of spacing.} \label{Fig3}
\end{center}
\end{figure}

For simplicity, we analyze first the absorption/emission based on the cross section of isolated antennas, although the interaction between different sites in an array can produce some well-understood corrections \cite{paper235}, as we discuss below. The absorption cross-section of a self-standing antenna reduces to $\sigma^{\rm abs}\approx(4\pi\omega/c){\rm Im}\{\alpha_\omega\}$ under normal incidence. We use this expression to obtain the radiated power per antenna within each plasmon resonance of frequency $\omega_j$ as $\approx (8\pi e^2/3\hbar^2c)A_j E_F L^2 E(\omega_j)$, assuming a Drude model for the conductivity $\sigma(\omega)=(\ii e^2E_F/\pi\hbar^2)/(\omega+\ii\gamma)$, which also predicts $\omega_j=(e/\hbar)\sqrt{B_j[2/(1+\varepsilon)]E_F/\pi L}$. This analytical model produces nearly indistinguishable results when compared with the spectral integral of the absorption calculated with the local-RPA response. The temperature dependence, which is fully contained in the distribution function $E(\omega)$ (see above), is analyzed in Fig.\ \ref{Fig3} using both models. Incidentally, the response also depends on temperature through the thermal smearing of the conduction electron band \cite{FP07_2}, although this is a small effect because we are in the $k_BT_1\ll E_F$ regime. The emission increases with temperature, and interestingly, the fraction of emission within the longitudinal, long-wavelength mode stays at a high level $>90\%$ up to temperatures above 700\,K. The addition of a silica substrate ($\epsilon=2$, included through Eq.\ (\ref{alpha}) and also taking into account the modification in the emission near an interface \cite{paper053}) only produces an overall increase in the emission, mainly towards the dielectric side [see Fig.\ \ref{Fig3}(a)]. We should emphasize again that this high level of efficiency relies on the poor absorption of graphene over a wide spectral range determined by $E(\omega)$, so that practical implementations of this idea should fight any residual absorption in this material or in the substrate.

The interaction between antennas arranged in periodic arrays produces corrections in the emission efficiency, as well as moderate shifts in the emission peaks, as illustrated in Fig.\ \ref{Fig3}(b), where we compare the absorption of a single antenna to simulations for extended antenna arrays using a dipole-based analytical model \cite{paper182}, in which we calculate the array absorptivity averaged over polar angles and light polarizations for the two principal azimuthal directions. Although the maximum absorptivity of a plane wave by a self-standing array is $50\%$  \cite{paper182}, averaging over polarizations and incidence directions limits this value to $\sim20\%$ for an optimized separation between the antennas [period $a=120\,$nm, see Fig.\ \ref{Fig3}(c), left scale]. Also, the absorption efficiency per antenna decreases due to this interaction at relatively small separations [Fig.\ \ref{Fig3}(b), right scale], although the integral of the main absorption peak over wavelengths yields a peak strength approaching the single island limit.

In conclusion, we have shown that electrically doped graphene can mediate the conversion of incident light into thermal radiation emitted at lower frequency via a suitable tailoring of the optical absorption profile. Our engineered structure allows achieving a concentration of the emission $>90\%$ within a single narrow resonance. More efficient devices should be realizable upon optimization of geometrical and material parameters, with the constraint that the emitting elements have to be thermally isolated in order to prevent diffusive cooling. The electrical tunability of graphene can be used to change both the emission frequency \cite{BSS14} and the absorption resonance. This could be achieved through a backgate that could simultaneously act as a mirror to reflect the backward emission. An interesting interplay takes place between the incident light power and the temperature reached by the structure before an equilibrium of the absorbed and emitted powers is achieved. A more efficient light extraction device could be designed by tunneling the graphene thermal emission through a resonant dielectric cavity placed in its vicinity, as suggested by a recent study of near-field radiative cooling \cite{GOP12}. This type of approach could also be exploited to guide the emission through a single output channel (e.g., a waveguide), in contrast to the broad angular emission distribution here studied. Our proposed concept of optical thermal conversion opens an unexplored approach towards the efficient generation of infrared light.

This work has been supported by the European Commission (Graphene Flagship CNECT-ICT-604391 and FP7-ICT-2013-613024-GRASP). A. M. acknowledges financial support from the Welch foundation through the J. Evans Attwell-Welch Postdoctoral Fellowship Program of the Smalley Institute of Rice University (Grant No. L-C-004).


\end{document}